\documentclass{ws-p8-50x6-00}

\begin{document}

\title{The Quest for Quantum Gravity: Testing Times for Theories?}

\author{N.E. Mavromatos}

\address{Theoretical Physics Group, Physics Department, 
King's College London, The Strand, London WC2R 2LS, United Kingdom}


\maketitle

\abstracts{I discuss some theoretical ideas concerning 
the representation of quantum gravity as a 
Lorentz-symmetry-violating `medium'
with non-trivial optical properties, which include 
a refractive index in `vacuo' 
and stochastic effects associated with a spread
in the arrival times of photons, growing linearly with the 
photon energy. Some of these properties
may be experimentally detectable 
in future satellite facilities (e.g. GLAST or AMS), using as probes 
light from distant astrophysical sources such as gamma ray bursters.
I also argue that such linear violations of Lorentz symmetry may not 
always be constrained by ultra-high-energy cosmic-ray data,
as seems to be the case with a specific (stringy) model of space-time foam.}

\begin{minipage}{4.2truein}
{\small Presented at the {\it Lake Louise Winter Institute 2000: From 
Particles to the Universe}, Lake Louise, Alberta, Canada, 20-26 February 2000.}
\end{minipage} 

\section{Introduction} 

The suggestion made by J.A. Wheeler~\cite{wheeler}
that space time may acquire a discrete foamy structure
at sub-Planckian scales has received considerable attention. 
The relevant works span a wide range of research fields, 
from phenomenological approaches~\cite{phenom} to theoretical modelling
of quantum gravity~\cite{qg} and/or string theory~\cite{string}.
The purpose of this talk is to focus on a recent scenario on the 
emergence of a foamy space-time structure in the context of string 
theory~\cite{emn},
and to discuss briefly its possible phenomenological 
consequences, especially in an astrophysical 
context using gamma-ray-bursters (GRB) 
as the relevant probes~\cite{aemns,efmmn}.

An important feature of most models of quantum space time 
foam is the breaking of Lorentz Invariance (LI) by quantum 
gravity effects. In such an approach LI is only an approximate
symmetry of the low-energy world.  In the context of the 
specific model of (stringy) space time foam proposed in \cite{emn}, 
the basic idea may be summarised as follows:
consider a closed-string state propagating in a 
$(D+1)$-dimensional space-time, which impacts
a very massive $D$(irichlet) particle embedded in this space-time.
In the modern view of string theory, $D$ particles must be included
in the consistent formulation of the ground-state vacuum
configuration.
We argue that the scattering of the closed-string state on the 
$D$ particle induces recoil of the latter, which
distorts the surrounding space-time 
in a stochastic manner. 

From a 
string-theory point of view,
the essential feature is the 
deviation from conformal invariance 
of the relevant world-sheet $\sigma$ model that
describes the recoil. This is compensated by
the introduction of a Liouville field~\cite{distler89}, which in
turn is identified with the target time
in the approach~\cite{emn} adopted here.

\section{Non-Trivial Optical Properties of Space-Time Foam} 

It has been pointed out in \cite{aemn},
based on some models of space-time foam in the context of 
Liouville strings, that the quantum-gravity `medium' may affect
the optical properties of the vacuum by inducing, among others, 
effects associated with a non-trivial refractive 
index. 

Similar phenomena have also been found~\cite{loopfoam} 
within the so-called `loop gravity'
approach to the dynamics of quantum space-time~\cite{ashtekar}.
In what follows we shall 
describe briefly the phenomenon by 
restricting ourselves, for definiteness, to a specific stringy model 
of quantum space time foam, in which one encounters
massive $D$-brane defects~\cite{emn}. 

In such a picture, the recoil of the 
massive space-time defect, during the scattering with the low-energy probe
(e.g. photon or neutrino), curves the surrounding space time, giving rise to 
a gravitational field of the form 
$G_{ij} \sim \eta_{ij} + {\cal O}\left(\frac{E}{M_Dc^2}\right)$,
where $c$ is the velocity of light in empty space,  
$E << M_D$ is the photon energy, and $M_D$ the gravitational 
scale of the defect. 
In string theory, $M_D = M_s/g_s$, where 
$M_s$ is the string scale, and $g_s << 1$ is the string coupling 
(assumed weak). One may identify $M_D$ with the Planck scale $10^{19}$ GeV 
in four space-time dimensions, or keep the $M_D$ as a phenomenological 
parameter to be constrained by observations~\cite{emn,efmmn}.

An important effect of such a distortion of space time 
is the appearance of an induced index of refraction: the effective
(group) velocity $v$ of photons in the quantum-gravitational 
`medium' depends in a way proportional to the energy of the 
particle probe:
\begin{equation} 
v=c\left(1 - {\cal O}\left(\frac{E}{M_Dc^2}\right)\right)
\label{refrindex}
\end{equation}
where the minus sign reflects the fact that there are {\it no} superluminal 
propagation in the $D$-brane recoil approach to space time foam.
This latter property has to be contrasted with some models in 
the loop gravity approach~\cite{loopfoam},
where superluminal  effects are present, leading to 
a dependence on the helicity of the photon state and thus 
characteristic birefringence effects. 
Notice that, since the space-time foam effects (\ref{refrindex}) 
are proportional to the energy of the particle probe, 
the phenomenon is quite distinct from conventional electromagnetic plasma 
effects, which attenuate with increasing energy. 
As emphasised in \cite{emn}, the effect (\ref{refrindex})
appears as a mean-field effect. Such an effect may be either severely 
constrained~\cite{mestres}
by ultra-high-energy ($10^{20}$ eV) 
cosmic-ray (UHECR) data~\cite{olinto} or
be viewed as implying that the standard assumptions on the maximum 
distance travelled by very-high-energy radiation have to be 
revised~\cite{kifune}. 

In addition to the above mean-field effects, 
there are {\it stochastic} fluctuations
about this mean value, which manifest themselves as light-cone fluctuations,
leading to a stochastic spread in the arrival times of photons
from a source at distance $L$
of the form~\cite{Ford}:
\begin{equation} 
\Delta t = \frac{\sqrt{<\sigma^2>-<\sigma_0>^2}}{L}
\label{stochastic}
\end{equation}
where $\sigma=\sum_{n=0}^{\infty} \sigma_n$, 
with $\sigma_n$ denoting the $n-th$ order term, with 
respect to an expansion in powers of the gravitational
fluctuations $h_{\mu\nu}$ about flat space time, of the
squared geodesic separation $2\sigma(x,x')$ between 
two space-time points $x$ and $x'$. 
The calculation for the recoil case, then, yields~\cite{emn}:
$\Delta t = {\cal O}\left(g_s\frac{EL}{M_Dc^3}\right)$,  
where the suppression by the extra power of $g_s$, as compared with 
the mean-field effect (\ref{refrindex}), is due to the fact that the effect
(\ref{stochastic}) represents quantum fluctuations about the mean value.  

Notice that the effect (\ref{stochastic}) is {\it not} associated with 
any modification of the dispersion relation of the particle probes,
but pertains strictly to fluctuations in the arrival 
times of photons~\cite{emn,Ford}. 
In fact, from its construction, the effect
is associated with the quantum uncertainties 
(notably probe-energy dependent~\cite{emn}) about the mean 
value of the recoil velocity along the incident direction, say $U_x$.
These are {\it non zero} even in the 
case of a vanishing mean-field $U_x$ that may 
characterize models of isotropic space time foam. In such isotropic scenaria
one averages $U_x$ over all possible directions, which 
leads to a vanishing refractive index effect (\ref{refrindex}). 
This in fact provides a possible counterargument
on recent claims~\cite{mestres} that violations of Lorentz symmetry 
which grow linearly with the particle energy 
may be incompatible with 
UHECR data. 
Incidentally, if one adopts the point of view
that UHECR may come from sources within a 50 Mpc radius from us,
as seems to have been suggested by the photon-pion production off the 
cosmic-background radiation~\cite{olinto}, 
then one would obtain a temporal dispersion of UHECR due to such effects 
over a period of $10^8$ sec. 

\section{Astro-particle Physics Phenomenology of Stringy Space-Time Foam} 

Although the effects (\ref{refrindex}) and 
(\ref{stochastic}) are tiny, however they are enhanced the further
the photon travels. Hence, 
data from distant cosmological sources, such as GRB, should provide 
a stringent constraint on these effects.

\begin{figure}[t]
\epsfxsize=3in
\bigskip
\centerline{\epsfbox{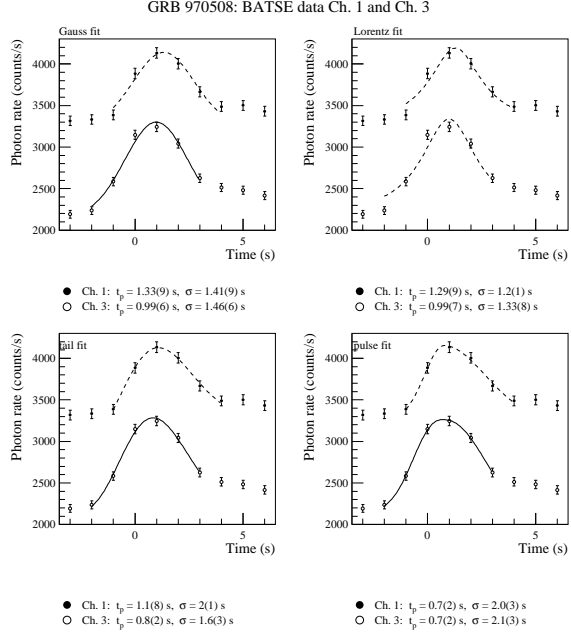}}
\caption{\it Time distribution 
of the number of photons 
observed by BATSE in Channels 1 and 3 for GRB~970508, compared 
with the following fitting functions: (a) Gaussian, (b)
Lorentzian, (c) `tail' function, and (d) `pulse' function.
We list below each panel the positions $t_p$ and widths $\sigma_p$
(with statistical errors) found for each peak in each fit.}
\label{figdat1}\end{figure}

\begin{figure}[t]
\epsfxsize=3in
\bigskip
\centerline{\epsfbox{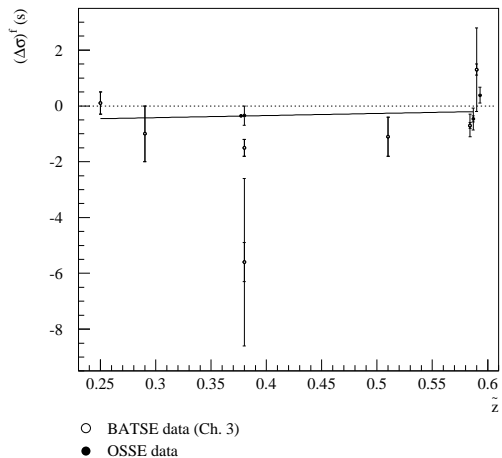}}
\caption{\it 
Values of the changes $(\Delta \sigma)^f$ in
the widths of the peaks fitted for each GRB studied
using BATSE and OSSE data, plotted versus
${\tilde z}=1-(1+z)^{-1/2}$, where $z$ is the
redshift. The indicated errors are the
statistical errors in the `pulse' fits provided by the fitting
routine,
combined with systematic error estimates obtained
by comparing the results obtained using the `tail' fitting
function. The values obtained by comparing OSSE with BATSE
Channel 3 data. The solid line is the best linear fit.}
\bigskip
\label{figregwidth}\end{figure}

We presented in~\cite{efmmn} a detailed 
analysis of the astrophysical data for 
a sample of 
GRB 
whose redshifts $z$ are known (see fig. \ref{figdat1}
for the data of a typical burst: GRB~970508).
We looked (without success) 
for a correlation with the redshift,
calculating a regression measure 
for the 
stochastic effect (\ref{stochastic}) (c.f. figure (\ref{figregwidth})), 
but also for the refractive 
index effect (\ref{refrindex}). 

We determined limits on
the quantum gravity scale $M_D$ 
by constraining the possible magnitudes of the slopes  in 
linear-regression analyses of the differences between the 
widths and arrival times
of pulses in different energy ranges from five GRBs with
measured redshifts,
as functions of the cosmically expanded redshift 
${\tilde z}$.
Using the current value for the Hubble expansion parameter,
$H_0 = 100 \cdot h_0$~km/s/Mpc, where $0.6 < h_0 < 0.8$,
we obtained 
$M_D \ge {\cal O}\left(10 ^{15}\right) \; {\rm GeV}$
on the possible quantum-gravity effects.

\section{Conclusions} 

We have discussed here some possible low-energy 
astrophysical probes of
quantum gravity, concentrating on
the possibility that the velocity of light might
depend on its frequency, i.e., the corresponding photon energy.
This idea is very speculative, and the model calculations that we have
reviewed require justification and refinement. However, we
feel that the suggestion is well motivated by the basic fact
that gravity abhors rigid bodies, and the
related intuition that the vacuum should exhibit back-reaction
effects and act as a non-trivial medium. 
These features have appeared in several approaches to quantum gravity,
including the canonical approach and ideas based on extra dimensions.

As could be expected, we have found no significant effect in the data
available on GRBs~\cite{efmmn}, 
either in the possible delay times of photons of
higher energies, or in the possible stochastic spreads of
velocities of photons with the same energy. 
However, it has been established that such probes may be
sensitive to scales approaching the Planck mass, if these
effects are linear in the photon energy.
We expect that the redshifts of many more GRBs will become
known in the near future.  

Future observations of higher-energy photons from GRBs~\cite{GLAST} 
would be
very valuable, since they would provide a longer
lever arm in the search for energy-dependent effects on photon
propagation. 
As emphasised in the text, such observations
are {\it essential} for the stochastic quantum gravity effect 
(\ref{stochastic}),  
and hence should be considered {\it complementary} 
to cosmic ray or other astrophysical data, which may not be 
relevant for constraining this effect.

\section*{Acknowledgments}

I wish to thank the organisers of 
the 2000 Lake Louise Winter Institute
for the warm hospitality and for providing a thought stimulating 
atmosphere during the conference.

\end{document}